\documentclass[twocolumn, trackchanges,]{aastex631}

\usepackage{multirow}

\begin{document}

\title{On the spin period distribution of millisecond pulsars}

\author[0000-0002-2187-4087]{Xiao-Jin Liu}
\affiliation{Advanced Institute of Natural Sciences, Beijing Normal University, Zhuhai 519087, China}
\affiliation{Department of Astronomy, Beijing Normal University, Beijing 100875, China}

\author[0000-0002-3309-415X]{Zhi-Qiang You}
\affiliation{Advanced Institute of Natural Sciences, Beijing Normal University, Zhuhai 519087, China}
\affiliation{Department of Astronomy, Beijing Normal University, Beijing 100875, China}
\affiliation{Henan Academy of Sciences, Zhengzhou 450046, Henan, China}

\author[0000-0001-7016-9934]{Zu-Cheng Chen}
\affiliation{Advanced Institute of Natural Sciences, Beijing Normal University, Zhuhai 519087, China}
\affiliation{Department of Astronomy, Beijing Normal University, Beijing 100875, China}

\author[0000-0002-0986-218X]{Shen-Shi Du}
\affiliation{School of Physics and Technology, Wuhan University, Wuhan 430072, China}
\affiliation{Advanced Institute of Natural Sciences, Beijing Normal University, Zhuhai 519087, China}

\author[0000-0001-9849-3656]{Ang Li}
\affiliation{Department of Astronomy, Xiamen University, Xiamen, Fujian 361005, China}

\author[0000-0001-7049-6468]{Xing-Jiang Zhu}
\affiliation{Advanced Institute of Natural Sciences, Beijing Normal University, Zhuhai 519087, China}
\correspondingauthor{Xing-Jiang Zhu}
\email{zhuxj@bnu.edu.cn}

\begin{abstract}
Spin period distribution provides important clues to understand the formation of millisecond pulsars (MSPs). 
To uncover the intrinsic period distribution, we analyze three samples of radio MSPs in the Galactic field and in globular clusters. 
The selection bias due to pulse broadening has been corrected but turns out to be negligible. 
We find that all the samples can be well described by a Weibull distribution of spin frequencies.
Considering MSPs in the Galactic field or in globular clusters, and in isolation or in binary systems, we find no significant difference in the spin distribution among these subpopulations.
Based on the current known population of MSPs, we find that sub-millisecond pulsars are unlikely to be discovered by the Square Kilometer Array, although up to $\sim10$ discoveries of pulsars that spin faster than the current record holder of $P=1.4$~ms are expected.

\end{abstract}


\section{Introduction} 
\label{sec:intro}

Pulsars are fast-spinning neutron stars whose electromagnetic radiation regularly sweeps over the Earth. 
Their strong gravitation, dense matter and diverse observational features at various wavelengths have made them favourable objects to study for a range of topics, including the search for nanohertz gravitational waves \citep{AAA+23, AAAB+23, RZS+23, XCG+23}, theories of gravity \citep{WH16, KSM+21}, neutron star equation of state \citep{HPY07}, ionised interstellar medium \citep{YMW17} and the evolution of binary systems \citep{TLK12}.

Currently, more than 3000 pulsars have been discovered, among which $\sim 560$ are millisecond pulsars (MSPs) \citep[see PSRCAT\footnote{\url{https://www.atnf.csiro.au/research/pulsar/psrcat/}},][]{MHT+05}. 
Both numbers will increase significantly in the next few years when new telescopes like the Five-hundred-meter Aperture Spherical radio Telescope (FAST) and MeerKAT complete their first rounds of pulsar surveys \citep{SLK+09, LWQ+18, HWW+21, PBS+23}. 
More importantly, when the square kilometer array (SKA) is used for pulsar search, the numbers are expected to further grow several folds \citep{SKS+09}, possibly finding a large fraction of all Milky-Way pulsars that are beaming toward the Earth \citep{KBK+15}.   

With such a large number of pulsars, statistical uncertainties due to small sample size become less significant and population analysis has become an effective way to study various population characteristics, including luminosity, birth rate, spatial \citep{Nar87, CC97, LFL+06} and kinematic distribution \citep{HLL+05, VIC17, Igo20}. 
It is also an important method to constrain braking index and study the decay of surface magnetic field, which plays a key role in pulsar evolution \citep{PB81, JK17, DPM22}.  

The intrinsic period distribution is a long-standing problem in pulsar population studies. 
A well-established period distribution is an essential input in population simulations \citep{BLR+14}, which are used to predict survey yields \citep[e.g.][]{KJv+10, NCB+15}, to estimate the total number of pulsars in the Galaxy, and, by comparing the simulated sample with survey results, to study other population characteristics \citep{FK06}.  

For MSPs, an accurate period distribution is critical to infer the value of the minimum spin period, which may impose significant constraints on theories of dense matter, since the maximum spin frequency (i.e., the Keplerian frequency) of pulsars depends sensitively on the equation of state governing their internal composition \citep{HZB+09, WEX+15}. 
It is also important in answering whether there are sub-millisecond pulsars \citep{PCD+98, BDM+01, LMB+13, LEP+15}, which are expected theoretically \citep[e.g.][]{FIP86,DXQ+09} but have not been discovered despite numerous search attempts \citep[e.g.,][]{EvB01, HML+04, DPF11}. 
In addition, joint analysis of period and period derivative distributions were used to study the spin-up history of MSPs \citep{PR72, GL79, TLK12, LYZ22} and the accretion physics during their low-mass X-ray phases \citep{Cha08, PTR+14, PHA17, CE23}. 
 
The period distribution of MSPs was firstly studied by \cite{CC97}. 
Using a sample of 22 MSPs, they studied a power-law model with a lower cut-off and found the minimum period $>0.65$~ms with 99 percent confidence. 
However, the power-law model was challenged by \cite{LEP+15}, which studied a sample of 56 MSPs discovered by various surveys that used the Parkes (Murriyang) radio telescope. 
Using maximum likelihood analysis, \cite{LEP+15} studied four two-parameter models and found that a log-normal distribution was favoured over the power-law model.
Using a sample of 337 radio MSPs, \cite{PHA17} found that, in the spin-frequency space, a Weibull model performed better than the log-normal distribution.

The studies of pulsar population are plagued with selection effects \citep{Lor11}, and so is the study of period distribution. 
Several factors can introduce period dependency into the pulsar detection rate. 
Notable ones include pulse dispersion and scattering, which increase the effective pulse width and thus the flux threshold to make a detection \citep[e.g.][]{DTW+85, CC97, Lor08}. 
The selection effects thus should be taken into account in the studies of period distribution.

In this paper, we study the period distribution of MSPs using the largest available sample from known pulsar surveys, which is more than four times the sample size of \cite{LEP+15}.
We have also developed a self-consistent Bayesian framework that accounts for selection effects in exploring the underlying period distribution. 
Compared with the grid-searching method used in \cite{LEP+15}, the Bayesian framework is more efficient in sampling high-dimensional likelihoods, enabling us to study more complex models and reduce the possibility of model misspecification. 
More importantly, the Bayesian method provides a natural way to incorporate selection effects into the analysis \citep[e.g.][]{HKD17, VGF+22} and allows robust model selection and parameter estimation.

The structure of the paper is as follows. 
We firstly describe the pulsar data and corresponding survey parameters in Section~\ref{sec:data}, then describe the Bayesian framework for parameter estimation and model comparison in Section~\ref{sec:Bayes}. 
After that, Section~\ref{sec:selection_effect} outlines the computation of the selection effects, while Section~\ref{sec:models} gives the period distribution models in consideration and Section~\ref{sec:results} presents our results. 
Finally, the results are discussed and concluded in Section~\ref{sec:summary}.  

\section{The Data} 
\label{sec:data}

Here, MSPs are defined as pulsars with a spin period, $P$, shorter than $20$~ms.
Our fiducial sample is Sample A, whose selection criteria are specified below.  

We only use radio pulsars and excluded pulsars discovered at other wavelengths, as the selection effects may vary significantly. 
We also exclude pulsars in globular clusters, as their spatial distribution differs from that of field pulsars (see Appendix~\ref{sec:ps}), making them unfit for the treatment of selection effects considered here.
Furthermore, we focus on pulsars discovered by surveys around 1400~MHz (L-band), avoiding the uncertainty introduced by extrapolating flux density measured at different frequencies. 

Since the computation of selection effects depends sensitively on survey parameters, we could only use surveys with well-defined parameters, i.e. the parameters should not change too many times over the survey lifetime. 
Using the Galactic MSP catalogue\footnote{\url{http://astro.phys.wvu.edu/GalacticMSPs/GalacticMSPs.txt}}, which collects both MSP parameters and corresponding discovery surveys, 46 different surveys were found. 
To make an efficient analysis, we set a yield threshold of 10 pulsars and 13 surveys satisfies the requirement. 
Since most radio MSPs were discovered in blind surveys, we thus exclude targeting searches of Fermi sources to maintain sample homogeneity. 
We are left with five surveys: the Parkes Multibeam Pulsar Survey (PMPS), High Time Resolution Universe Southern survey (HTRUS), Pulsar Arecibo L-band Feed Array (PALFA) survey, Commensal Radio Astronomy FasT survey (CRAFTS) and Galactic Plane Pulsar Snapshot (GPPS) survey. 
The five surveys and corresponding parameters are listed in Table~\ref{tab:survey}. 
We note that both CRAFTS\footnote{\url{http://groups.bao.ac.cn/ism/CRAFTS/CRAFTS}} and GPPS\footnote{\url{http://zmtt.bao.ac.cn/GPPS/GPPSnewPSR.html}} are on-going projects and have a number of MSPs that are not yet included in either the Galactic MSP catalogue or PSRCAT.
We thus query the corresponding project webpages and add these new discoveries. 

\begin{figure}
    \centering
    \includegraphics[trim={0.2cm 0.3cm 1.4cm 0.8cm}, clip, width=\columnwidth]{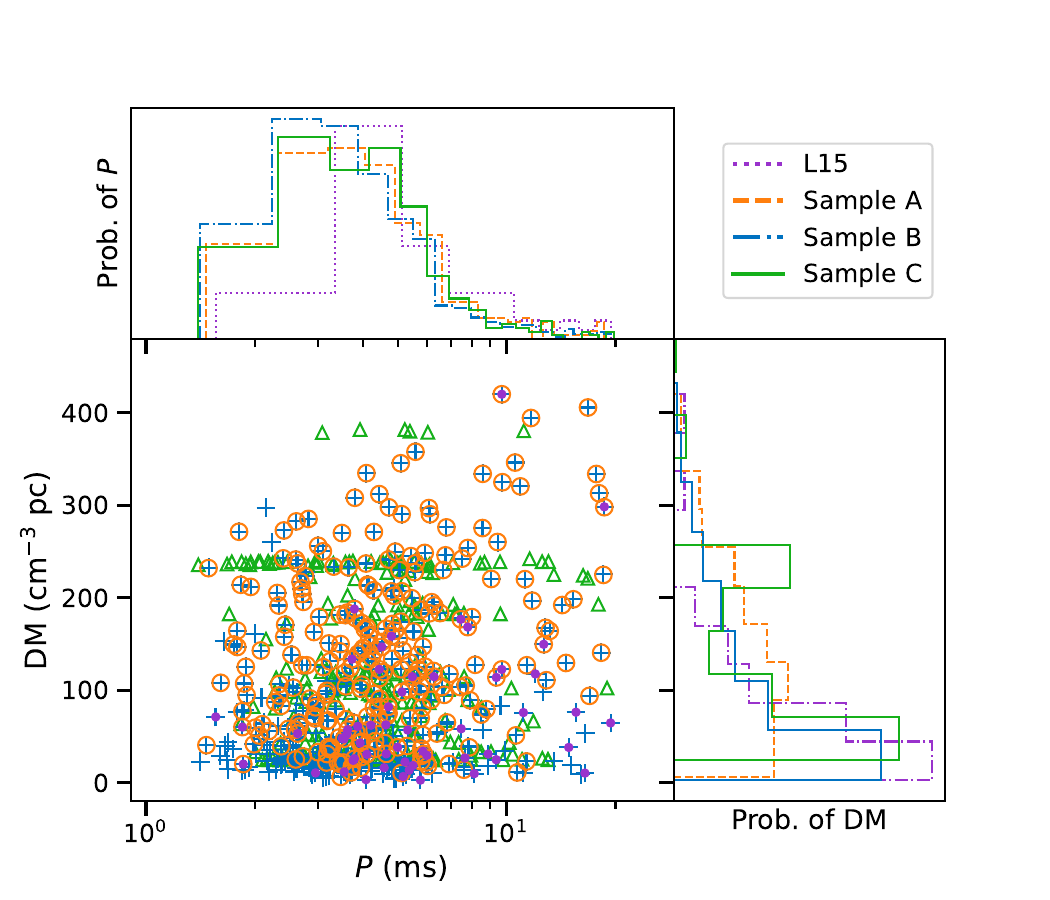}
    \caption{The period and DM distribution of the MSPs in \cite{LEP+15} (L15, purple dots, 58 MSPs), Sample A (yellow circles, 260 MSPs), Sample B (blue plus symbol, 473 MSPs) and Sample C (green triangles, 273 MSPs).
    The normalized histogram of both period and DM are also plotted with the corresponding color. The pulsar samples are available at \url{https://github.com/pulsar-xliu/Field_MSP_sample}.}
    \label{fig:psrhist}
\end{figure}

In total, there are 260 MSPs in Sample A, which is more than four times the sample size of \cite{LEP+15}.
The normalized period histogram of Sample A and that of \cite{LEP+15} are shown in Fig~\ref{fig:psrhist}.
It is clear that Sample~A has a higher portion of faster MSPs than \cite{LEP+15}. 
Since dispersion measure (DM) also affects the inference of period distribution (see Section~\ref{sec:selection_effect} for details), Fig.~\ref{fig:psrhist} shows the DM value of both samples. 
Compared with the \cite{LEP+15} sample, Sample~A has a higher portion of large DM pulsars. 

For a more complete picture, we also compile a list of all the radio MSPs in the Galactic field (Sample B, 473 MSPs) and a list of all MSPs in globular clusters\footnote{\url{https://www3.mpifr-bonn.mpg.de/staff/pfreire/GCpsr.html}} (GCs, Sample C, 273 MSPs).
Compared with Sample A, Sample B has the fastest Galactic field pulsar (J0952$-$0607, 1.41~ms, \citealt{BPH+17}) and slightly more MSPs below 4~ms, while Sample~C has the fastest pulsar J1748$-$2446ad (1.396~ms, \citealt{HRS+06}). Both samples are also shown in Fig.~\ref{fig:psrhist}. 

\begin{table}
    \footnotesize
    \centering
        \caption{The survey parameters and corresponding number of MSPs ($N_{\rm MSP}$) used in Sample A. 
        The meaning of parameters are explained in Section~\ref{sec:Smin}. 
        Notes: (a) Two different receivers were used in both PALFA and CRAFTS, but most PALFA MSPs were discovered using Mock spectrometer system, while most CRAFTS MSPs were discovered by 19-beam receiver. 
        We picked out these MSPs and used the survey settings in corresponding receiver phases. 
        (b) As $G$ of multibeam receivers decreased from the core  to outer beams, typical values were used here for simplicity.
        $G$ is in units of Jy$^{-1}$K.
        (c) The $\beta$ factor is estimated using the method given by~\protect\cite{JA98}. 
        (d) As $t_{\rm int}$ could depend on $l$ and $b$, the correct $t_{\rm int}$ was given to each MSP according to their $l$ and $b$. 
        References in the table are: (1) \protect\cite{MLC+01}, (2) \protect\cite{KJv+10}, (3) \protect\cite{LBH+15}, (4) \protect\cite{LWQ+18}, (5) \protect\cite{HWW+21}.}
\begin{tabular}{cccccc}
\hline\hline
Surveys$^{\rm (a)}$  & PMPS	& HTRUS  & PALFA & CRAFTS & GPPS \\
    \hline
Telescope & Parkes & Parkes & Arecibo & FAST & FAST	 \\
Ref.  & (1)  & (2) 	& (3)	& (4)	& (5)	 \\
$N_{\rm MSP}$ 	& 23  & 36 	& 30 	& 36  & 135  \\
$G^{\rm (b)}$ & 0.70 & 0.70 & 8.2 & 16  & 16	\\
$T_{\rm sys}$ (K) 	& 21 & 23 & 24	& 20 & 18	\\
$t_{\rm samp}$ ($\mu$s) & 250 & 64 	& 65.5	& 49.15 & 49.15  \\
$f$ (MHz) 	& 1374 	& 1352 & 1375 & 1250 & 1250   \\
$B$ (MHz) 	& 288   & 340  & 322.6	& 400 	& 400 \\
$\Delta f$ (MHz) & 3 & 0.391 & 0.336 & 0.122 & 0.244  \\
$N_{\rm chan}$ 	& 96 & 870 	& 960	& 4096 	& 2048 \\
$n_{\rm pol}$ 	& 2  & 2  & 2	& 2  & 2 \\
$\beta^{\rm (c)}$ 	& 1.25 	& 1.16 	& 1.0	& 1.0 & 1.0   \\
\hline
 & 	& 270 	& 268 &  &     \\
$t_{\rm int}^{\rm (d)}$ (s) & 2100	& 540 & 180	& 15	& 300	\\
 &	& 4200 	& 	& 	&  \\
\hline
\end{tabular}
\label{tab:survey}
\end{table}

\section{The Bayesian method of parameter estimation and model comparison} \label{sec:Bayes}

Given a sample of MSP period $\{d\}$ and an underlying period distribution model M which has a set of parameters, $\Lambda$, Bayesian inference \citep[e.g.][]{TT19} constrains the model parameters through $P(\Lambda | d, {\rm M}) = \pi(\Lambda|{\rm M})L(d|\Lambda)/Z$, where $\pi(\Lambda|{\rm M})$ is the prior probability density function of the parameters, $L(d|\Lambda)$ is likelihood and $Z$ is model evidence. 
The likelihood is given by 
\begin{equation}
\label{intact_lhd}
    L(d|\Lambda) = \int L(d|P) \pi(P|\Lambda){\rm d}P = \pi(d|\Lambda),
\end{equation}
where $P$ denotes the period predicted by the underlying distribution and $\pi(P|\Lambda)$ is hyper-prior. 
Note that in the second equality of Eq.~\ref{intact_lhd}, we have used $L(d|P)=\delta(d-P)$, which is an excellent approximation in practice, as the measurement of pulse period is usually very precise.

Given selection effects $\mathbb{C}$, however, parts of expected data are filtered out by selection effects, leading to bias in both parameter estimation and model selection. 
Deriving the biased likelihood, $L(d|\Lambda, \mathbb{C})$, is thus a key step in Bayesian analysis. 
To obtain  $L(d|\Lambda, \mathbb{C})$, further using Bayes theorem and Eq.~\ref{intact_lhd}, then \citep[e.g.][]{HDK+19, MFG19}
\begin{equation}
    L(d|\Lambda, \mathbb{C}) 
    = \frac{p(\mathbb{C}|d,\Lambda)\pi(d|\Lambda)}{\int p(\mathbb{C}|P, \Lambda)\pi(P|\Lambda){\rm d}P},
\end{equation}
where $p(\mathbb{C}|d, \Lambda)$ is the selection effect for data $\{{d}\}$ and $p(\mathbb{C}|P, \Lambda)$ is the selection effect for a general data set. 
For the observed data set $\{{d}\}$, $p(\mathbb{C}|d,\Lambda) = 1$ according to definition, while for a general set of data, a detailed analysis of the selection effect is required and given in Section~\ref{sec:selection_effect}. 
Note that the selection effect usually has no direct dependence on population parameters, thus $p(\mathbb{C}|P,\Lambda) = p(\mathbb{C}|P)$. 
When the data are from different surveys, the total likelihood becomes 
\begin{equation}
    \label{eqn:Ltotal}
    L_{\rm total}(d|\Lambda, \mathbb{C})=\prod_i \frac{\pi(d_i|\Lambda)}{\int p(\mathbb{C}_i|P)\pi(P|\Lambda){\rm d}P},
\end{equation}
where $i$ is the index of $i-$th survey. 

By sampling over the parameter space of the likelihood, one can estimate both model parameters and evidence. 
For two models, M$_1$ and M$_2$, the natural-log-based Bayes factor is usually defined as 
$\ln {{\rm BF}^{{\rm M}_{1}}_{{\rm M}_{2}}} = \ln Z_{{\rm M}_{1}} - \ln Z_{{\rm M}_{2}}$,
where $Z_{{\rm M}_{1}}$ and $Z_{{\rm M}_{2}}$ are the evidence for each model. 
A positive log Bayes factor indicates preference of M$_{1}$ over M$_{2}$ \citep[e.g.][]{Jef61}.

\section{The selection effect} \label{sec:selection_effect}

For pulsar surveys, the major selection effect is caused by the limited survey sensitivity. The sensitivity depends on the survey parameters and scales as $(W/(P-W))^{1/2}$, where $W$ is
effective pulse width and $P$ is the pulse period. 
The effect is thus period-dependent and affects the inference of underlying period distribution. 
Following \cite{LEP+15}, the effect can be modelled in two steps: first deriving the detection rate of a given MSP, $\mathbb{D}_{\rm smear}$, then estimating the selection effect, $p_{\rm smear}$, by fitting $\mathbb{D}_{\rm smear}$ with a function of $P$ and DM. 
Specifically, $\mathbb{D}_{\rm smear}$ is given by 
\begin{equation}
    \label{Eqn:pSdetect}
    \mathbb{D}_{\rm smear}(P, {\rm DM}) = \frac{{\int_{S_{\rm min}}^\infty} p(S) {\rm d}S}{{\int_{S_0}^\infty} p(S) {\rm d}S},
\end{equation}
where $p(S)$ is the probability density function of flux density $S$, $S_{\rm min}$ is survey sensitivity, and $S_0$ is the limit of $S_{\rm min}$ at large $P$ and zero DM.
The effect is obtained by fitting $\mathbb{D}_{\rm smear}$ with 
\begin{equation}
  \label{eqn:psmear}
   p_{\rm smear} = \Big[1 - \exp{\Big(-\frac{P}{\alpha}}\Big)\Big]\times\exp{\Big(-\frac{\rm DM^2}{2\beta^2}\Big)}. 
\end{equation}
Since $S_{\rm min}$ varies from survey to survey, each survey has a set $\alpha$ and $\beta$. 
For readability, we give the fitting results of $\alpha$ and $\beta$ for the five surveys in Table~\ref{tab:psmear}, and show the computation details of $\mathbb{D}_{\rm smear}$ in Appendix~\ref{appendix:getSE}. 

\begin{table}
    \footnotesize
    \centering
    \caption{The coefficients of $p_{\rm smear}$ for the surveys in Sample A.}
    \begin{tabular}{cccccc}
    \hline\hline
    & \multicolumn{5}{c}{Survey} \\
    \cline{2-6}
       Parameter  & PMPS & HTRUS & PALFA & CRAFTS & GPPS \\
       \hline
       $\alpha$ (ms) & 3.50 & 0.94 & 1.00 & 0.62 & 0.90 \\
       $\beta$ (cm$^{-3}$~pc)& 192 & 600 & 567 & 1036 & 492 \\
       \hline
       \label{tab:psmear}
    \end{tabular}
\end{table}

\section{Modelling the period distribution} 
\label{sec:models}

Since the underlying period distribution is unknown, we consider ten different models to minimize the possibility of model misspecification. 
The models are mainly motivated by the observed distribution and can be classified into three categories as described below.

\subsection{Single-component models}

Following \cite{LEP+15}, we use normal, log-normal, and $\Gamma$ distribution to model the period distribution of MSPs. The normal distribution is denoted by $\mathcal{N}(\mu, \sigma)$,
\begin{equation}
\mathcal{N}(\mu,\sigma)
=\frac{1}{\sigma \sqrt{2 \pi}} \exp \left[-\frac{1}{2}\left(\frac{P-\mu}{\sigma}\right)^{2}\right],
\end{equation}
where $\mu$ is the mean and $\sigma$ is the width. 
The log-normal distribution is defined as
\begin{equation}
 {\rm log}\mathcal{N}(\mu, \sigma)=\frac{1}{\sqrt{2\pi}P\sigma}\exp\Big[-\frac{(\ln P - \mu)^2}{2 \sigma^2}\Big],
\end{equation}
where $\mu$ and $\sigma$ are the mean and width, respectively. 
The $\Gamma$ distribution is defined as  
\begin{equation}
    \Gamma(k, \theta) = \frac{1}{\Gamma(k)\theta^k}P^{k-1}\exp\Big[-\frac{P}{\theta}\Big],
\end{equation}
where $\Gamma(k)$ is the Gamma function and $k$ is shape parameter and $\theta$ is scale. 
We also consider a Weibull distribution, which reads as 
\begin{equation}
W(k, \theta)=\frac{k}{\theta}\left(\frac{P}{\theta}\right)^{k-1} \exp\Big[-\left(\frac{P}{\theta}\right)^k\Big],
\end{equation}
where both scale $\theta > 0$ and shape $k > 0$. A Weibull distribution of spin frequency as considered in \cite{PHA17} is equivalent to the following 
\begin{equation}
    W_\nu(k, \theta) = \frac{1}{P^2}\frac{k}{\theta}\Big(\frac{1}{P\theta}\Big)^{k-1} \exp\Big[-\Big(\frac{1}{P\theta}\Big)^k\Big],
\end{equation}
where the subscript indicates that the corresponding distribution in the frequency ($\nu$) space is Weibull. The priors of all the parameters are listed in Table~\ref{tab:PDF}. 

The models above are normalized before being applied for analysis. 
For the normalization, we use 20~ms (the sample selection criterion) as the upper boundary and use a fixed lower boundary of 0.6~ms, which is the Keplerian period limit computed from $f_{\rm K} = 1.15(M/{\rm M}_\odot)^{1/2}(R/10~{\rm km})^{-3/2}$~kHz \citep{HZB+09} using $M=2~{\rm M}_\odot$ and $R=10$~km. 

\subsection{Lower cut-off models}

Theoretically, rotating neutron stars should have a maximum spin frequency, or equivalently, a minimum spin period. 
However, the exact value of the minimum period is unclear and it is interesting to search for evidence of a minimum spin period in the spin distribution of MSPs. 
Therefore, we consider models with a lower cut-off, $P_{\rm min}$, which is set as a free parameter. 

The first cut-off model is modified from the normal distribution and given by $\mathcal{N}_{\rm cut} = \mathbb{H}(P-P_{\rm min}) \mathcal{N}(\mu, \sigma)$, where $\mathbb{H}(P-P_{\rm min})$ is the Heaviside function. 
Similarly, the second cut-off model is a modified version of the log-normal distribution and reads as ${\rm log}\mathcal{N}_{\rm cut} = \mathbb{H}(P-P_{\rm min}) {\rm log}\mathcal{N}(\mu, \sigma)$, while the third model is an adaption of $\Gamma(k, \theta)$: $\Gamma_{\rm cut} = \mathbb{H}(P-P_{\rm min}) \Gamma(K, \theta)$.
Finally, we also consider the once favoured power-law model, which also has a lower cut-off and reads as $L_{\rm Pow} = \mathbb{H}(P-P_{\rm min}) P^\gamma$. 
 
In all the cut-off models, we use a uniform prior, $U(0.1, 2)$ ms, for $P_{\rm min}$. 
The priors of other parameters are also uniform and given in Table~\ref{tab:PDF}. 
Slightly different from the single-component models, all the cut-off models are normalized in the range of [$P_{\rm min}$, 20~ms].

\subsection{Two-component models}

The population of MSPs can be divided into different groups, e.g., isolated and binary MSPs, it is reasonable to speculate if they follow different period distributions. 
To study this possibility, we consider three two-component models.

The first model consists of two normal distributions: $L_{(\mathcal{N}+\mathcal{N})} = w \mathcal{N}(\mu_1, \sigma_1) + (1-w) \mathcal{N}(\mu_2, \sigma_2)$, where $w$ is weight and we set $\mu_1 < \mu_2$ to avoid confusion between the two components. 
Similarly, the second model is a mixture of normal and log-normal model, 
$L_{(\mathcal{N}+{\rm log}\mathcal{N})} = w \mathcal{N}(\mu_1, \sigma_1) + (1 - w) {\rm log}\mathcal{N}(\mu_2, \sigma_2)$, while the third model is a mixture of the normal and Gamma models: $L_{(\mathcal{N} + \Gamma)} = w \mathcal{N}(\mu, \sigma) + (1-w) \Gamma(k, \theta)$.
For all the three models above, a uniform prior, $U(0, 1)$, is given to $w$. 
The priors for other parameters are also uniform and given in Table~\ref{tab:PDF}.
The models are normalized in the range of [0.6, 20]~ms, the same as that of single-component models. 

\renewcommand{\arraystretch}{1.1}
\begin{table*}
    \caption{Summary of period distribution models used in this study. Columns are model names, parameters, priors, {\it maximum a posteriori} values (with 1-$\sigma$ error) and natural-log Bayes factors (with respect to that of the normal model).}
    \centering
    \small
    \begin{tabular}{ccccccccccccc}
    \hline\hline
    & & & \multicolumn{4}{c}{Sample A} & & \multicolumn{2}{c}{Sample B} & & \multicolumn{2}{c}{Sample C}\\
    \cline{4-7} \cline{9-10} \cline{12-13} 
    & & & \multicolumn{2}{c}{SE Modelling} & \multicolumn{2}{c}{No SE Modelling} & & \multicolumn{2}{c}{No SE Modelling} & & \multicolumn{2}{c}{No SE Modelling}\\ 
    \cline{1-7} \cline{9-10} \cline{12-13} 
   Model & Par. &  Prior & Posterior$_1$ & $\ln$BF$_1$ & Posterior$_2$ & $\ln$BF$_2$ & & Posterior$_3$ & $\ln$BF$_3$ & & Posterior$_4$ & $\ln$BF$_4$\\
   \cline{1-7} \cline{9-10} \cline{12-13}
     \multirow{2}{*}{$\mathcal{N}$} & $\mu$ [ms] & $U(0.1, 5)$ &  $0.90^{+0.84}_{-0.57}$ & \multirow{2}{*}{0} & $1.84^{+0.94}_{-1.03}$ & \multirow{2}{*}{0} & & $0.96^{+0.75}_{-0.58}$ & \multirow{2}{*}{0} & & $2.94^{+0.61}_{-0.83}$ & \multirow{2}{*}{0}\\ 
     & $\sigma$ [ms] & $U(0.1, 8)$ & $5.37^{+0.34}_{-0.38}$ & & $5.22^{+0.50}_{-0.49}$ & &  & $5.23^{+0.29}_{-0.34}$ & & & $4.22^{+0.46}_{-0.36}$ \\
   \cline{2-7} \cline{9-10}  \cline{12-13}
     \multirow{2}{*}{{\rm log}$\mathcal{N}$} & $\mu$ & $U(0.1, 3)$ & $1.48^{+0.04}_{-0.04}$ & \multirow{2}{*}{36.9} & $1.51^{+0.03}_{-0.03}$ &  \multirow{2}{*}{45.8} & &  $1.43^{+0.03}_{-0.03}$ & \multirow{2}{*}{89.7} & & $1.48^{+0.03}_{-0.03}$ & \multirow{2}{*}{54.2} \\ 
    & $\sigma$ & $U(0.1, 2)$ & $0.56^{+0.03}_{-0.03}$ & & $0.54^{+0.03}_{-0.02}$ & &  & $0.54^{+0.02}_{-0.02}$ & & & $0.50^{+0.02}_{-0.02}$\\ 
    \cline{2-7} \cline{9-10}  \cline{12-13}
    \multirow{2}{*}{$\Gamma$} & $k$ & $U(0.1, 5)$ & $2.96^{+0.32}_{-0.30}$ & \multirow{2}{*}{21.5} &  $3.27^{+0.30}_{-0.29}$ & \multirow{2}{*}{28.7} & &  $3.18^{+0.22}_{-0.22}$ & \multirow{2}{*}{53.4} & & $3.89^{+0.34}_{-0.33}$ & \multirow{2}{*}{36.6}\\
     & $\theta$ [ms] & $U(0.1, 5)$ & $1.71^{+0.20}_{-0.16}$ & & $1.60^{+0.17}_{-0.15}$ & &  & $1.53^{+0.12}_{-0.11}$ & & & $1.28^{+0.13}_{-0.11}$\\
   \cline{2-7} \cline{9-10}  \cline{12-13}
    \multirow{2}{*}{$W$} & $k$ & $U(0.1, 3)$ & $1.53^{+0.09}_{-0.09}$ & \multirow{2}{*}{10.0} &  $1.64^{+0.09}_{-0.09}$ & \multirow{2}{*}{13.1} & &  $1.58^{+0.06}_{-0.06}$ & \multirow{2}{*}{24.5} & & $1.76^{+0.08}_{-0.08}$ & \multirow{2}{*}{15.9}\\
    & $\theta$ [ms] & $U(0.1, 8)$ & $5.41^{+0.26}_{-0.26}$ & & $5.78^{+0.24}_{-0.24}$ & &  & $5.33^{+0.17}_{-0.17}$ & & & $5.52^{+0.21}_{-0.21}$\\
    \cline{2-7} \cline{9-10}  \cline{12-13}
    \multirow{2}{*}{$W_\nu$} & $k$ & $U(0.1, 3)$ & $1.98^{+0.12}_{-0.12}$ & \multirow{2}{*}{\bf 40.5} &  $1.97^{+0.11}_{-0.11}$ & \multirow{2}{*}{\bf 49.2} & &  $2.09^{+0.09}_{-0.09}$ & \multirow{2}{*}{\bf 110.0} & & $2.20^{+0.12}_{-0.11}$ & \multirow{2}{*}{\bf 57.3}\\
    & $\theta$ [ms$^{-1}$] & $U(0.1, 8)$ & $0.29^{+0.01}_{-0.01}$ & & $0.28^{+0.01}_{-0.01}$ & &  & $0.30^{+0.01}_{-0.01}$ & & & $0.28^{+0.01}_{-0.01}$\\
   \cline{1-13}       
    \multirow{3}{*}{$\mathcal{N}_{\rm cut}$} & $P_{\rm min}$ [ms] & $U(0.1, 2)$ & $1.45^{+0.01}_{-0.03}$  & & $1.45^{+0.01}_{-0.03}$ & &  & $1.41^{+0.01}_{-0.01}$ & & & $1.38^{+0.01}_{-0.03}$ & \multirow{3}{*}{30.0}\\
  & $\mu$ [ms] & $U(0, 5)$ & $0.34^{+0.49}_{-0.25}$ & 21.8 & $0.39^{+0.56}_{-0.28}$ &  31.0 & &  $0.27^{+0.26}_{-0.12}$ & 60.7 & & $0.68^{+0.77}_{-0.42}$\\
 & $\sigma$ [ms] & $U(0.1, 8)$ & $5.30^{+0.26}_{-0.26}$ & & $5.41^{+0.27}_{-0.27}$ & &  & $5.12^{+0.18}_{-0.19}$ & & & $4.89^{+0.28}_{-0.31}$\\    
    \cline{2-7} \cline{9-10}  \cline{12-13}
    \multirow{3}{*}{\rm log$\mathcal{N}_{\rm cut}$} & $P_{\rm min}$ [ms] & $U(0.1, 2)$ & $1.43^{+0.03}_{-0.09}$ & & $1.43^{+0.03}_{-0.08}$ & &  & $1.40^{+0.01}_{-0.03}$  & & & $1.29^{+0.09}_{-0.57}$ & \multirow{3}{*}{\bf 55.9}\\
    & $\mu$ & $U(0.1, 3)$ & $1.44^{+0.04}_{-0.05}$ & {\bf 41.0} &  $1.47^{+0.04}_{-0.04}$ &  {\bf 50.1} & &  $1.38^{+0.03}_{-0.03}$ & {\bf 101.7} & & $1.47^{+0.03}_{-0.03}$\\
    & $\sigma$ & $U(0.1, 2)$ & $0.60^{+0.04}_{-0.04}$  & & $0.59^{+0.04}_{-0.03}$ & &  & $0.59^{+0.03}_{-0.03}$ & & & $0.51^{+0.03}_{-0.03}$\\   
    \cline{2-7} \cline{9-10}  \cline{12-13}
    \multirow{3}{*}{$\Gamma_{\rm cut}$} & $P_{\rm min}$ [ms] & $U(0.1, 2)$ & $1.45^{+0.01}_{-0.04}$ & & $1.45^{+0.01}_{-0.03}$ & &  & $1.41^{+0.01}_{-0.01}$ & & & $1.37^{+0.02}_{-0.05}$ & \multirow{3}{*}{43.5}\\
    & $k$ & $U(0.1, 5)$ & $2.01^{+0.36}_{-0.37}$ & 32.4 & $2.24^{+0.37}_{-0.36}$ & 41.0 & &  $2.02^{+0.27}_{-0.26}$ & 83.6 & & $3.10^{+0.40}_{-0.38}$ \\
    & $\theta$ [ms] & $U(0.1, 5)$ & $2.29^{+0.41}_{-0.31}$ & & $2.15^{+0.35}_{-0.28}$ & &  & $2.17^{+0.26}_{-0.23}$ & & & $1.54^{+0.19}_{-0.17}$\\
    \cline{2-7} \cline{9-10}  \cline{12-13}
    \multirow{2}{*}{$L_{\rm Pow}$} & $P_{\rm min}$ [ms] & $U(0.1, 2)$ & $1.46^{+0.01}_{-0.01}$ & \multirow{2}{*}{$2.0$} & $1.46^{+0.01}_{-0.01}$ & \multirow{2}{*}{$6.6$} & &  $1.41^{+0.00}_{-0.01}$ & \multirow{2}{*}{22.7} & & $1.39^{+0.00}_{-0.01}$ & \multirow{2}{*}{$-14.2$}\\    
    & $\gamma$ & $U(-3, 0)$ & $-1.42^{+0.08}_{-0.09}$ & & $-1.33^{+0.09}_{-0.08}$ & &  & $-1.42^{+0.06}_{-0.06}$ & & & $-1.32^{+0.08}_{-0.08}$\\
   \cline{1-13} 
     \multirow{5}{*}{${(\mathcal{N}+\mathcal{N})}$} & $w$ & $U(0, 1)$ & $0.72^{+0.07}_{-0.08}$ & & $0.72^{+0.07}_{-0.08}$ & &  & $0.72^{+0.06}_{-0.06}$ & & & $0.76^{+0.06}_{-0.07}$ & \multirow{5}{*}{49.3}\\
     & $\mu_1$ [ms] & $U(0.1, 5)$ & $3.88^{+0.14}_{-0.15}$  & & $4.01^{+0.13}_{-0.13}$ & &  & $3.69^{+0.09}_{-0.09}$ & & & $4.02^{+0.12}_{-0.12}$ \\
     & $\sigma_1$ [ms] & $U(0.1, 4)$ & $1.57^{+0.17}_{-0.14}$ & 36.3 & $1.50^{+0.14}_{-0.13}$ & 43.9 & &  $1.36^{+0.09}_{-0.09}$ & 88.3 & & $1.44^{+0.13}_{-0.11}$\\
    & $\mu_2$ [ms] & $U(0.1, 15)$ & $7.18^{+2.48}_{-2.16}$ & & $7.89^{+2.26}_{-2.21}$ & &  & $7.26^{+1.72}_{-1.94}$ & & & $7.48^{+2.04}_{-1.98}$\\
    & $\sigma_2$ [ms] & $U(0.1, 10)$ & $7.12^{+1.65}_{-1.49}$ & & $6.73^{+1.75}_{-1.42}$ & &  & $6.32^{+1.46}_{-1.08}$ & & & $6.03^{+1.78}_{-1.25}$\\    
    \cline{2-7} \cline{9-10}  \cline{12-13}
     \multirow{5}{*}{${(\mathcal{N} + {\rm log}\mathcal{N})}$} & $w$ & $U(0, 1)$ & $0.16^{+0.17}_{-0.12}$  & & $0.08^{+0.15}_{-0.07}$ & &  & $0.48^{+0.12}_{-0.13}$ & & & $0.47^{+0.19}_{-0.30}$ & \multirow{5}{*}{51.5}\\ 
    & $\mu_1$ [ms] & $U(0.1, 8)$ & $4.59^{+0.73}_{-0.25}$  & & $4.87^{+1.05}_{-0.35}$ & &  & $3.42^{+0.17}_{-0.22}$ & & & $3.92^{+0.28}_{-0.25}$\\
    & $\sigma_1$ [ms] & $U(0.1, 3)$ & $1.22^{+0.45}_{-0.35}$ & 36.7 & $1.20^{+0.77}_{-0.45}$ &  42.9 & &  $1.08^{+0.15}_{-0.18}$ & 94.2 & & $1.25^{+0.18}_{-0.21}$\\
    & $\mu_2$ & $U(0.1, 3)$ & $1.46^{+0.04}_{-0.04}$  & & $1.49^{+0.04}_{-0.04}$ & &  & $1.69^{+0.15}_{-0.10}$ & & & $1.63^{+0.24}_{-0.14}$\\
    & $\sigma_2$ & $U(0.1, 2)$ & $0.61^{+0.08}_{-0.05}$ & & $0.57^{+0.06}_{-0.04}$ & &  & $0.62^{+0.06}_{-0.05}$ & & & $0.57^{+0.09}_{-0.06}$\\    
    \cline{2-7} \cline{9-10}  \cline{12-13}
    \multirow{5}{*}{${\rm (\mathcal{N}+\Gamma)}$} & $w$ & $U(0, 1)$ & $0.71^{+0.07}_{-0.08}$ & & $0.70^{+0.06}_{-0.08}$ & &  & $0.70^{+0.05}_{-0.06}$ & & & $0.73^{+0.07}_{-0.08}$ & \multirow{5}{*}{51.3}\\
    & $\mu$ [ms] & $U(0.1, 5)$ & $3.85^{+0.15}_{-0.15}$ & & $3.91^{+0.14}_{-0.14}$ & &  & $3.58^{+0.10}_{-0.10}$ & & & $3.93^{+0.13}_{-0.14}$\\
     & $\sigma$ [ms] & $U(0.1, 5)$ & $1.44^{+0.15}_{-0.14}$ & 31.8 & $1.42^{+0.14}_{-0.13}$ & 45.5 & &  $1.27^{+0.10}_{-0.09}$ & 92.8 & & $1.35^{+0.12}_{-0.11}$\\
    & $k$ & $U(0.1, 5)$ & $2.91^{+0.94}_{-0.65}$  & & $3.02^{+0.93}_{-0.67}$ & &  & $2.98^{+0.75}_{-0.55}$ & & & $3.22^{+0.90}_{-0.70}$\\ 
    & $\theta$ [ms] & $U(0.1, 5)$ & $3.10^{+0.99}_{-0.70}$  & & $3.03^{+1.01}_{-0.68}$ & &  & $2.78^{+0.74}_{-0.51}$ & & & $2.50^{+0.85}_{-0.52}$\\
    \hline
    \end{tabular}
    \label{tab:PDF}
\end{table*}
\renewcommand{\arraystretch}{1.0}

\section{Results} 
\label{sec:results}

Using the Bayesian framework and the \textsc{ptemcee} sampler\footnote{\url{https://github.com/willvousden/ptemcee}} \citep{VFM21}, we obtain the posterior distributions of the parameters and corresponding model Bayesian evidence, which are shown in Table~\ref{tab:PDF}. 
In the analysis of Sample A, we consider both the case with and without selection effect modelling.
We also analyze Samples B and C, which contain all the radio MSPs in the Galactic field and globular clusters, respectively. 
Due to the complex discovery history in Sample B and the significantly different spatial distribution in Sample C, we only considered the case without selection effect and show the results in Table~\ref{tab:PDF}.

Table~\ref{tab:PDF} shows that, no matter the selection effect is modelled or not, both the $W_\nu$ and {\tt log$\mathcal{N}_{\rm cut}$} model are strongly favoured by Sample A, although {\tt log$\mathcal{N}_{\rm cut}$} is marginally ($\ln {\rm BF}\le 0.9$) better. 
While using Sample B, $W_\nu$ is strongly favoured over {\tt log}$\mathcal{N}_{\rm cut}$ with $\ln {\rm BF}=8.3$. 
In Sample C, $W_\nu$ remains the best model, but the {\tt log$\mathcal{N}_{\rm cut}$} model cannot be ruled out as the relative $\ln {\rm BF}$ is only 1.4. 
In all cases, two-component models are strongly disfavoured.

\begin{figure}
    \centering    
    \includegraphics[trim={0.2cm 0.4cm 0.6cm 0.8cm}, clip, width=\columnwidth]{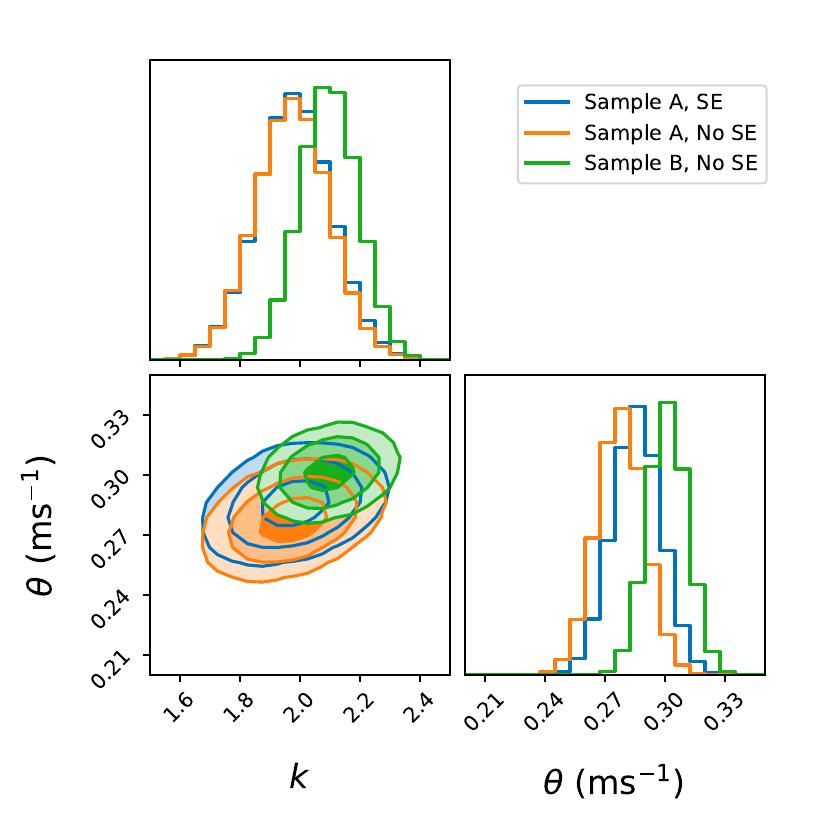}    
    \caption{Posterior distributions of the  {\tt $W_\nu$} model. 
    For Sample A, the results with selection effect modelling are in blue, while those without the effect are in yellow. 
    The results for Sample B without the effect are in green.}
    \label{fig:PDF_2p}
\end{figure}

\begin{figure*}
    \centering
    \includegraphics[trim={0.1cm 0.5cm 0.1cm 0.1cm}, clip, width=0.8\textwidth]{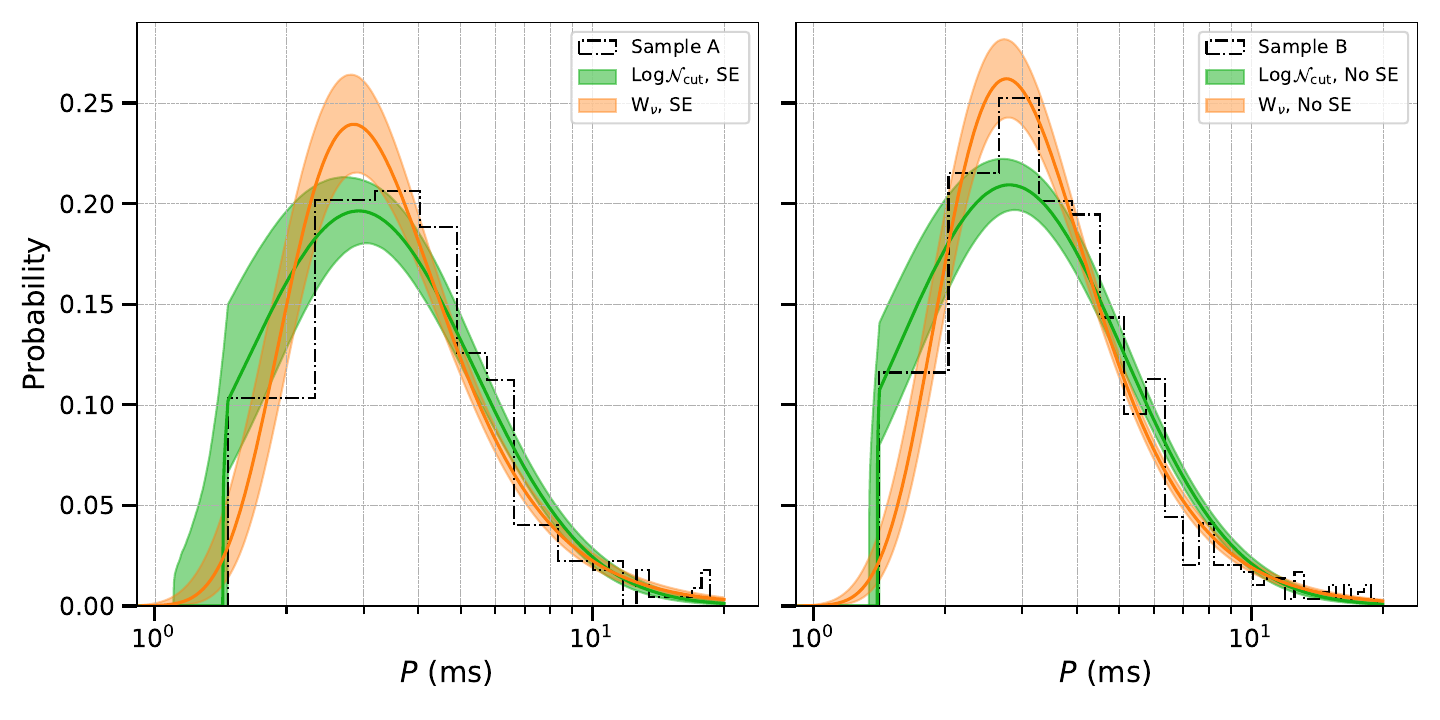}
    \caption{Comparison of the posterior predictive distribution between the {\tt log}$\mathcal{N}_{\rm cut}$ and $W_\nu$ model. 
    The shaded area indicates the 90\% confidence regions, while the dashed lines show the histograms of the samples.
    {\bf Left}: The distributions for Sample A considering the selection effect due to pulse broadening.
    {\bf Right}: Results for Sample B without modelling the selection effect.}
    \label{fig:ppd}
\end{figure*}

For posteriors, there are only slight changes for the parameters of $W_\nu$ for all the cases, as exemplified in Fig.~\ref{fig:PDF_2p} for Samples A and B.
The value of both parameters (shape $k\sim2$ and scale $\theta \sim 0.29~{\rm ms}^{-1} =290~{\rm Hz}$) are consistent with that of \cite{PHA17}.
The fitness of the $W_\nu$ and {\tt log$\mathcal{N}_{\rm cut}$} model to Samples A and B are shown by the posterior predictive distributions in Fig.~\ref{fig:ppd}.

Using the posterior predictive distribution of the $W_\nu$ model, we computed the fraction of sub-millisecond pulsars ($f_{\rm sub}$) and the fraction of fast pulsars, i.e. pulsars with $P\le 1.396$~ms, $f_{\le\rm 1.396}$.
Both fractions are shown in Table~\ref{tab:fsub}.

We also tested the robustness of our method by simulating mock samples using the $W_\nu$ model (see Appendix~\ref{apx:robustness} for details). 
Our analysis successfully recover the injected parameters and the $\ln {\rm BF}$ decisively favour the true model over others.

\begin{table}[]
    \centering
    \caption{ 
The fraction of sub-millisecond and fast pulsars ($P\le1.396$~ms) for different cases (considering different samples and the inclusion of selection effects or not) using the $W_\nu$ model. 
Also listed are the corresponding expected discoveries of sub-millisecond and fast pulsars, assuming a low and high yield of the SKA pulsar survey.}
    \begin{tabular}{cccccc}
    \hline\hline
     Par. & Sample & SE & Fraction & $N_{\rm low}$ & $N_{\rm high}$\\
    \hline
     & A & Y & $9.5\times10^{-6}$ & 0.01 & 0.06\\
    \multirow{2}{*}{$f_{\rm sub}$} & A & N & $4.8\times10^{-6}$ & 0.007 & 0.03\\
     & B & N & $4.3\times10^{-6}$ & 0.006 & 0.03\\
     & C & N & $7.3\times10^{-8}$ & 0.0001 & 0.0004\\
    \hline
     & A & Y & $2.6\times10^{-3}$ & 3.9 & 15.5\\
    \multirow{2}{*}{$f_{\le 1.396}$} & A & N & $1.8\times10^{-3}$ & 2.7 & 10.7\\
     & B & N & $2.1\times10^{-3}$ & 3.2 & 12.9\\
     & C & N & $3.8\times10^{-4}$ & 0.6 & 2.3\\
    \hline
    \end{tabular}
    \label{tab:fsub}
\end{table}

\section{Discussion and Conclusions}
\label{sec:summary}

\subsection{Robustness of the results}
The results can be affected by several factors, including selection effects and the choice of MSP samples. 
We discuss these factors below.

First, using Sample A, one sees that for both {\tt log}$\mathcal{N}_{\rm cut}$ and $W_\nu$ model, the selection effect due to pulse broadening has very limited impacts on both the posteriors (Fig.~\ref{fig:PDF_2p}) and the log Bayes factors (Table~\ref{tab:PDF}).
The small impacts of selection effects, especially for samples discovered with modern instruments, were also noticed by \cite{LEP+15}.
The small selection effects thus warrant us to extend the analysis from Sample A to Samples B and C.

Second, in addition to pulse broadening, beaming also introduces period-dependent bias to the observed MSP sample \citep[e.g.][]{LM88, EC89, Lor08}, as it affects the probability of a pulsar beaming toward the Earth hence its detectability. 
For MSPs, there is no empirical relation of beaming fraction; however, the fraction is believed to be between 0.4 and 1, and possibly close to unity \citep{KXL+98, LMB+13}. 
More importantly, faster pulsars usually have a higher beaming fraction and thus are less affected by beaming. 
Therefore, the impact of beaming on the constraint of $P_{\rm min}$ should be small.

Additionally, binary motion modulates the observed spin frequency and leads to period-dependent loss of signal-to-noise ratios \citep{Lor08}. 
This impact has been largely overcome in recent surveys, especially for wide-orbit binaries, by the extensive use of acceleration search.
For the impact on compact binaries with a small spin period, it remains to be investigated. 
However, no significant difference was found in the period distribution between the isolated and binary systems, for both \cite{LEP+15} and our sample (see Appendix~\ref{sec:other_sel} for further details).

Finally, the parameter posteriors are relatively stable with respect to the sample in use, but the Bayes factors are clearly affected by the sample size.
To see this more clearly, we also analyze the \cite{LEP+15} sample for the case with and without considering selection effects in their paper.
In both cases, the {\tt log$\mathcal{N}_{\rm cut}$} model is marginally better ($\ln {\rm BF}\sim 2$) than the $W_\nu$ model. 
Taking this into account and considering the large size of Sample B, we thus tend to believe, for radio MSPs, $W_\nu$ is more likely to be the underlying period distribution\footnote{We find no evidence of a minimum period cutoff while adopting the Weibull distribution.} than the other models considered here.

\subsection{Sub-millisecond pulsars and equation of state}

Irrespective of the sample and selection effect, the fraction of sub-millisecond pulsars obtained for the $W_\nu$ model remains $< 10^{-5}$.
The small $f_{\rm sub}$ is consistent with the lack of fast pulsars close to the break-up limit, a puzzle noticed in both nuclear- and accretion-powered MSPs \citep{CMM+03, Pat10}, which are free from the selection effect \citep{HRS+06} but have a much smaller sample size.

Since the SKA will likely find thousands of new MSPs \citep{SKS+09, KBK+15}, we also compute the expected number of sub-millisecond pulsars and fast pulsars. 
Given the yield uncertainty, we assume a low case of 1500 MSPs discoveries and a high case of 6000 discoveries by the SKA. 
In either case, the discovery of sub-millisecond pulsars is unlikely, with an expectation $< 0.1$, while the chance to discover fast pulsars looks higher, with expectations ranging from $\sim1$ to around 10 discoveries.

The lack of pulsars faster than $\sim 1$~ms has attracted great attention and several models have been proposed, including spin-down via gravitational radiation at extremely small spin periods \citep[e.g.][]{Cha08}, and the failure to spin-up to the break-up limit during the recycling phase  \citep[e.g.][]{BC17, EA21, CE23}. 

\begin{figure}
    \centering    
    \includegraphics[width=\columnwidth]{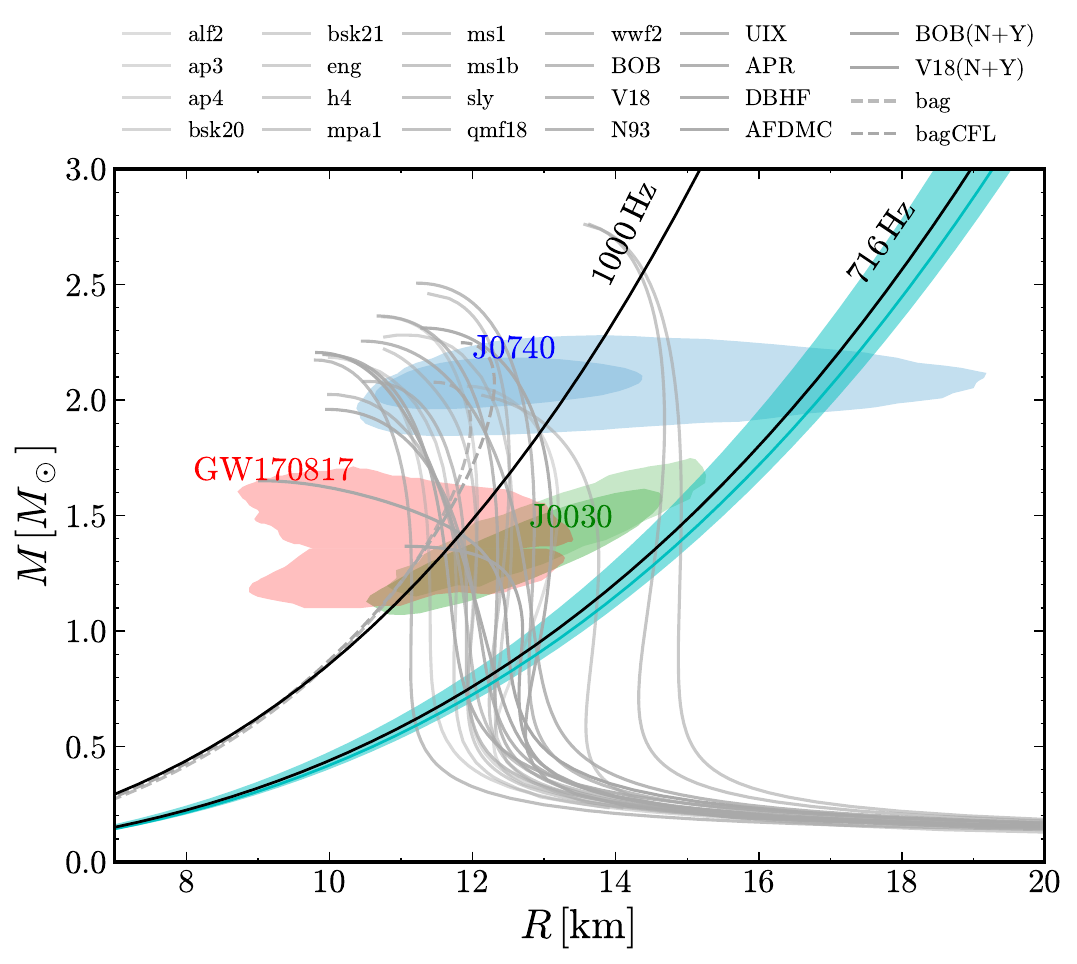}
    \caption{Mass-radius relations of neutron stars (with or without phase transition in cores; grey solid curves) and quark stars (in non-superfluid or superfluid states; grey dashed curves) for different equations of state \citep[][and references therein]{ZZL18, ZZL18_2,LZZ+20}. 
    The observational constrains are from GW170817 \citep{AAA+18EOS} and PSRs~J0030+0451 \citep{MLD+19, RWB+19} and J0740+6620 \citep{MLD+21, RWR+21}. 
    The black lines are the lower boundaries of the mass-radius space, using the current maximum spin rate (716~Hz, \citealt{HRS+06}) and an assumed maximum rate of 1000~Hz.
    The range for $P_{\rm min}=1.43^{+0.03}_{-0.09}$ from the {\tt log$\mathcal{N}_{\rm cut}$} model is also indicated near the 716-Hz curve.}
    \label{fig:MR}
\end{figure}

The lack of sub-millisecond pulsars will have an important impact on the study of dense matter equations of state: a practical equation of state must yield Kepler frequencies not less than the observed spin frequencies \citep{WEX+15}. 
The possible discovery of sub-millisecond pulsars has long been seen as a strong evidence of the existence of quark stars \citep{Gle90}.
However, as shown in Fig.~\ref{fig:MR}, the result of $P_{\rm min}\ge1$~ms, i.e. $f_{\rm max}\le 1000$~Hz, does not allow for discrimination between different models of equation of state, or prove the existence of quark stars in this way.

\subsection{Do different groups of MSPs follow the same period distribution?}
\label{sec:sub_groups}
Usually, population studies treat GC and field pulsars separately, partly due to the apparent difference in spatial distribution and formation history or environments. 
Their spin periods, however, seem to be well described by the $W_\nu$ model with comparable parameters ($k\sim2$ and $\theta\sim0.3$~ms$^{-1}$), although for GC pulsars, Sample C has a weaker preference for $W_\nu$ over {\tt log}$\mathcal{N}_{\rm cut}$, probably due to the smaller sample size.

\begin{figure}
    \centering
    \includegraphics[width=\columnwidth]{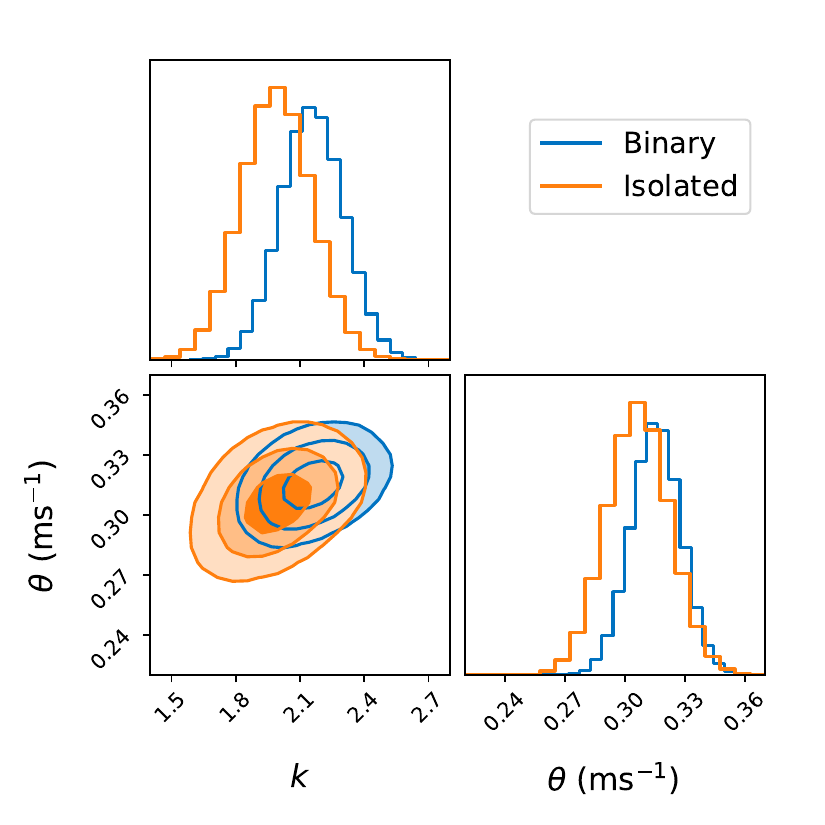}
    \caption{The posteriors of the isolated and binary radio MSPs in the Galactic field using the $W_\nu$ model.}
    \label{fig:iso_binary}
\end{figure}

Another interesting issue is whether isolated and binary pulsars follow the same period distribution.
We test the hypothesis with the isolated ($N=148$) and binary ($N=200$) radio MSPs in the Galactic field (taken from PSRCAT V1.70). Using the best-fit model, $W_\nu$, both samples give consistent posteriors within $1\sigma$ uncertainty, see Fig.~\ref{fig:iso_binary}.

\subsection{Conclusions}
Analyzing the spin period distribution of known MSPs, we find that a Weibull model best describes both MSPs from the selected surveys and all the radio MSPs in the Galactic field, with the best-fitting parameters of $k=2.09^{+0.09}_{-0.09}$ and $\theta=0.30^{+0.01}_{-0.01}~{\rm ms}^{-1}$, in agreement with \cite{PHA17}. 
Moreover, we find that the spin periods of MSPs in globular clusters also follow a Weibull distribution with consistent parameters.
We find no statistical support for a minimum spin period cutoff in the spin distribution of MSPs.
For the MSPs in the Galactic field, our analysis shows that both the isolated and binary pulsars follow the same parent period distribution. 
We also incorporate selection effects due to pulse broadening into the Bayesian analysis, and find that they are negligible.

Based on the Weibull model as inferred for the known MSP population, we find that the SKA will likely find a few pulsars faster than 1.396~ms (PSR J1748$-$2446ad). 
However, the fraction of sub-millisecond pulsars is $<10^{-5}$, which means an extremely low chance for the SKA to discover such pulsars.
The lack of sub-millisecond pulsars suggests that observed pulsar spins might not be an effective way to discriminate unusual equations of state of neutron stars.

\begin{acknowledgments}
We thank Dr. Lei Zhang for useful discussions. 
XJL is supported by China Postdoctoral Science Foundation (No.~2022M710428). 
ZQY is supported by the National Natural Science Foundation of China under Grant No.~12305059.
ZCC is supported by the National Natural Science Foundation of China (Grant No.~12247176 and No.~12247112) and China Postdoctoral Science Foundation (No.~2022M710429). 
AL is supported by the National Natural Science Foundation of China (Grant No.~12273028).
XJZ is supported by the National Natural Science Foundation of China (Grant No.~12203004).
\end{acknowledgments}

\software{Jupyter \citep{KRP+16}, Numpy \citep{HM20}, Matplotlib \citep{HUN07}, Astropy \citep{ART+13, APS+18, APL+22}, \textsc{ptemcee} \citep{VFM21}.}

\appendix
\section{Computing the selection effect due to pulse broadening} \label{appendix:getSE}

\subsection{The probability density function of flux density}
\label{sec:ps}

To compute the selection effect given by Eqn~\ref{sec:selection_effect}, one needs $p(S)$, which can be determined by
\begin{equation}
    \label{pS_los}
    p(S) = \int p(S|D) p(D) {\rm d}D,
\end{equation}
where $D$ is the distance to pulsar. 
Below, we give procedures to obtain $p(S|D)$ and $p(D)$. 

For observations at 1.4~GHz, $p(S|D)$ can be obtained using the probability density function of luminosity ($L$), which follows a lognormal distribution with a mean of $-1.1$ and a scatter of 0.9 \citep{FK06}. 
Since $L=SD^2$, we have 
\begin{equation}
    \label{eqn:pSD}
    p(S|D) \propto \Big|\frac{\partial L}{\partial S}\Big|p(L) \propto \frac{1}{S} \exp\Bigg[-\frac{\big(\log_{10}S +2\log_{10}D + 1.1 \big)^2}{2\times0.9^2}\Bigg],
\end{equation}
where $S$ and $D$ are in units of mJy and kpc, respectively. 

Given Galactic longitude ($l$) and latitude ($b$), the term $p(D)$ is \citep{LFL+06, VWC+12}
\begin{equation}
\label{GF_D_dist}
    p(D) \propto R^{1.9} D^2 \exp\Bigg[\frac{-|D\sin b|}{0.5\, \rm kpc}\Bigg]\exp\Bigg[-5 \frac{|R-R_0|}{R_0}\Bigg],
\end{equation}
where 
$R = \sqrt{R_0^2 + (D\cos b)^2 - 2R_0 D\cos b\cos l}$, 
and $R_0$ is the Solar distance to the Galactic centre.  
We use $R_0 = 8.2$~kpc \citep{GAA+20}.  

Numerically integrating Eqn.~\ref{pS_los} gives the $p(S)$ for a given set of $(l, b)$ or line-of-sight. 
Note that the adoption of Eqn.~\ref{eqn:pSD} restricts us to use MSPs discovered by surveys at $\sim$1.4~GHz.

\subsection{The survey sensitivity and detection rate}
\label{sec:Smin}

Another key ingredient in computing the selection effect is $S_{\rm min}$. 
Following \cite{DTW+85, LK12}, $S_{\rm min}$ can be estimated by 
\begin{equation}
    \label{Eqn:radiometer}
    S_{\rm min} = \frac{{\rm (S/N)}\beta T_{\rm sys}}{G\sqrt{n_{\rm p} t_{\rm int}B}}\sqrt{\frac{W_{\rm e}}{P- W_{\rm e}}},
\end{equation}
where S/N is detection threshold, $\beta$ is a factor describting digitization performance \citep[e.g.][]{JA98}, $T_{\rm sys}$ is system temperature, $G$ is telescope gain, $n_{\rm p}$ is the number of polarization, $t_{\rm int}$ is integration time, $B$ is receiver bandwidth, $W_{\rm e}$ is the effective pulse width and $P$ is pulse period. 
The value of all these parameters (except $T_{\rm sys}$ and $W_{\rm e}$) is listed in Table~\ref{tab:survey}. 
Below, we give the prescriptions to obtain $T_{\rm sys}$ and $W_{\rm e}$. 

The system temperature is given by $T_{\rm sys}=T_{\rm rec} + T_{\rm sky}$, where $T_{\rm rec}$ and $T_{\rm sky}$ are receiver and sky temperature, respectively. 
We computed $T_{\rm sky}$ using the Haslam sky model \citep{HKS+81} implemented in the \textsc{PyGDSM} package\footnote{\url{https://github.com/telegraphic/pygdsm}} \citep{PRI21}, which uses for extrapolation a spectral index of $-2.85$ between 400~MHz and a few GHz \citep{DBC+19}.  

The effective pulse width is given by 
\begin{equation}
    W_{\rm e} = \sqrt{W_{\rm i}^2 + t_{\rm samp}^2 + \bigg(\frac{\rm DM}{{\rm DM}_0}t_{\rm samp}\bigg)^2 + t_{\rm scat}^2},
\end{equation}
where $W_{\rm i}$ is the intrinsic pulse width, $t_{\rm samp}$ is sampling time, ${\rm DM}_0$ is the DM value that smears the pulse in one bin to $t_{\rm samp}$ and $t_{\rm scat}$ is scattering time. 
For $W_{\rm i}$, we use an 8 per cent duty cycle, i.e. $W_{\rm i}/P=0.08$, which is the median value of the MSPs in our sample. 
${\rm DM}_0$ is given by
\begin{equation}
   {\rm DM}_0=N_{\rm chan}\Big(\frac{t_{\rm samp}}{\rm s}\Big)\Big(\frac{f}{\rm MHz}\Big)^3\Big(\frac{\rm MHz}{8299 B}\Big), 
\end{equation}
where $N_{\rm chan}$ is the number of frequency channel. 
For $t_{\rm scat}$, we use the result from \cite{COR02} and ignore its scatter  
\begin{equation}
    \log_{10}\Big(\frac{t_{\rm scat}}{\mu{\rm s}}\Big) = -3.59 + 0.129\log_{10} {\rm DM} + 1.02 \big(\log_{10}{\rm DM}\big)^2 \\
    -4.4\log_{10}\Big(\frac{f}{\rm GHz}\Big). 
\end{equation}

Note that Eqn.~\ref{Eqn:radiometer} is prone to overestimating the flux density at small period for large DM values \citep{CC97, LBH+15}, as $S_{\rm min}$ increases rapidly when $W_{\rm e}$ approaches $P$. 
This feature will bias the estimation of selection effect, as it will reduce the value of the numerator in Eqn.~\ref{Eqn:pSdetect}, hence $\mathbb{D}_{\rm smear}$. 
A more appropriate, though complicated, way to estimate $S_{\rm min}$ was given by \cite{CC97}. 
Since most MSPs in Sample A have relatively small DM thus small degree of overestimation, we use Eqn.~\ref{Eqn:radiometer} for clarity.

\begin{figure}
    \centering
    \includegraphics[trim={5cm 0.2cm 4.2cm 1.5cm}, clip, width=0.45\columnwidth]{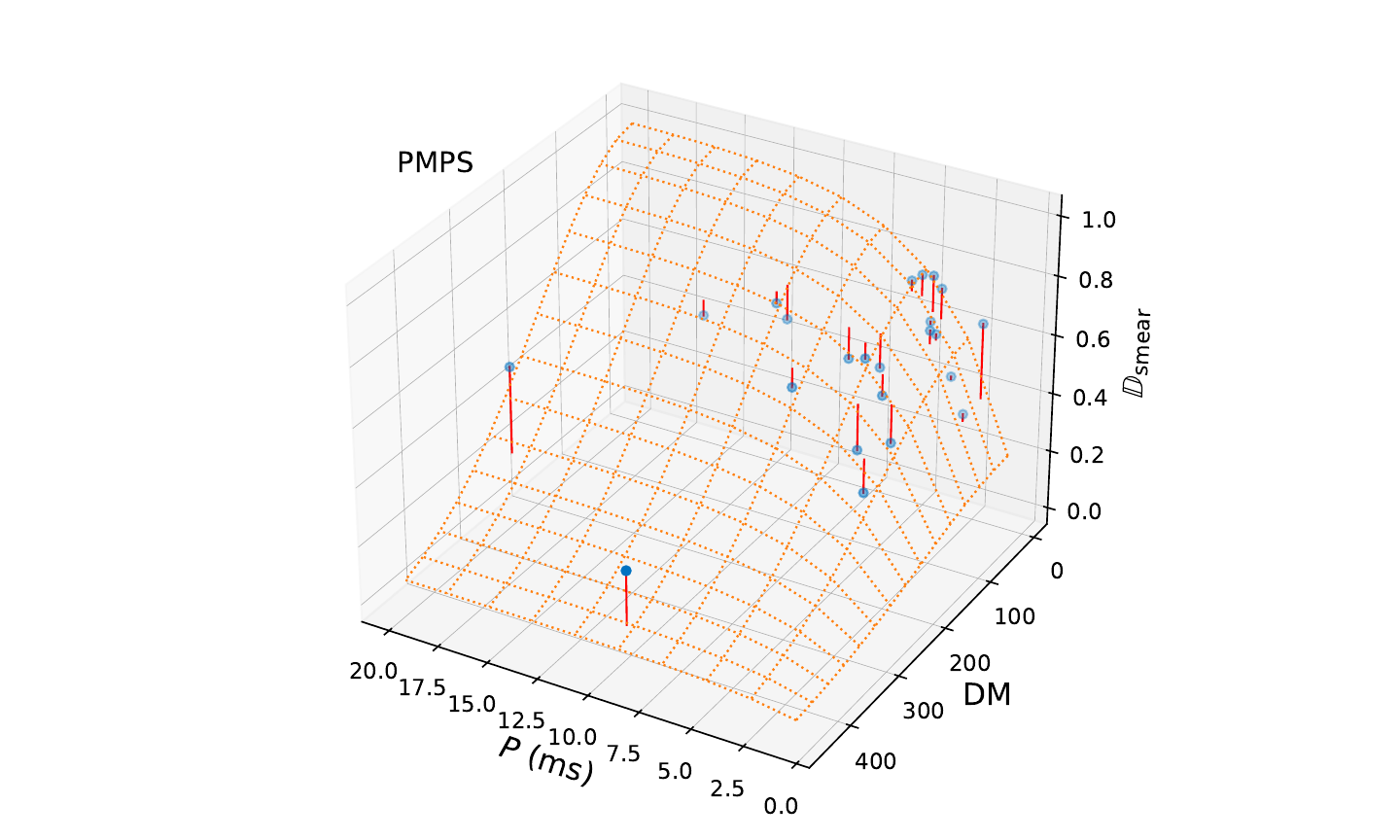}
    \includegraphics[trim={5cm 0.2cm 4.2cm 1.5cm}, clip, width=0.45\columnwidth]{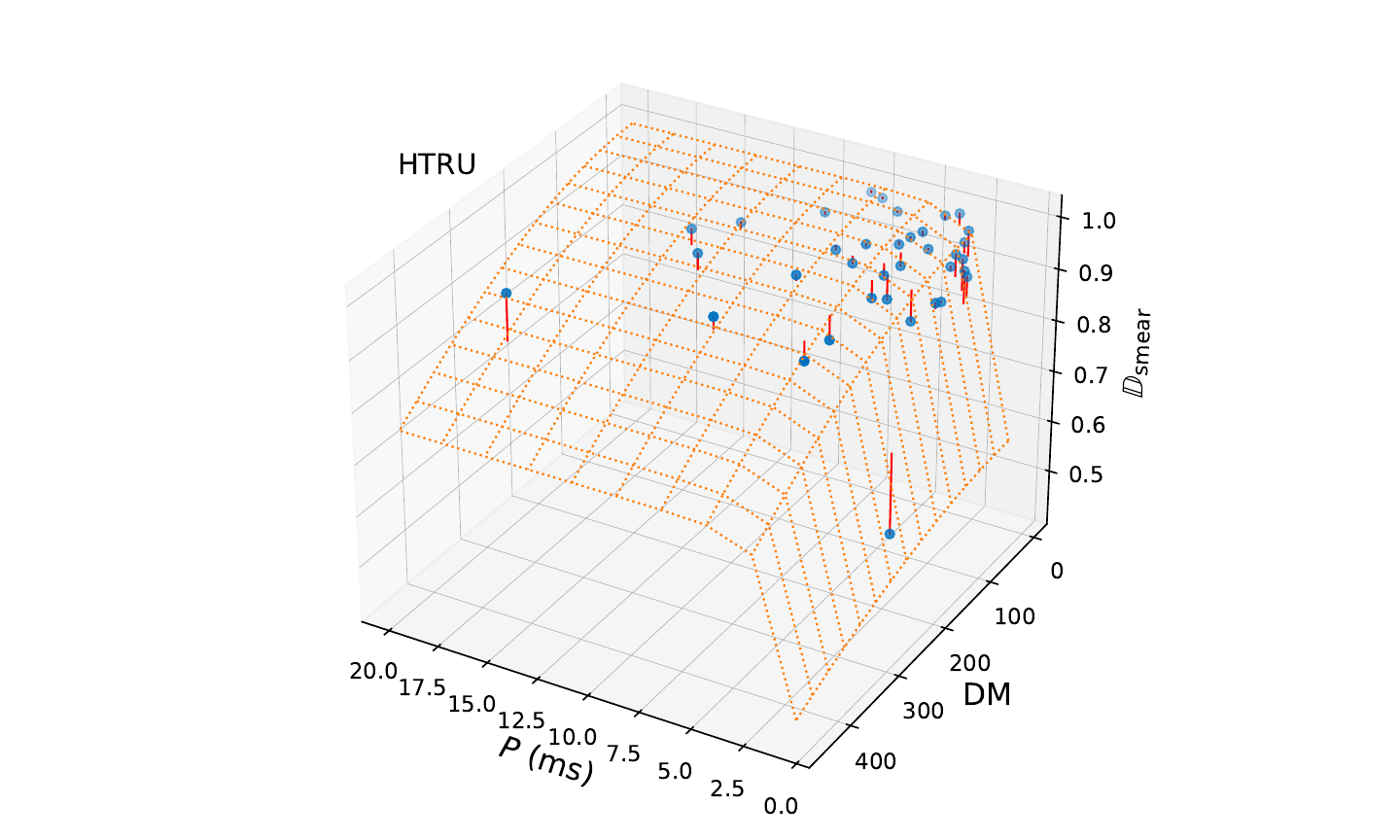}
    \includegraphics[trim={5cm 0.2cm 4.2cm 1.5cm}, clip, width=0.45\columnwidth]{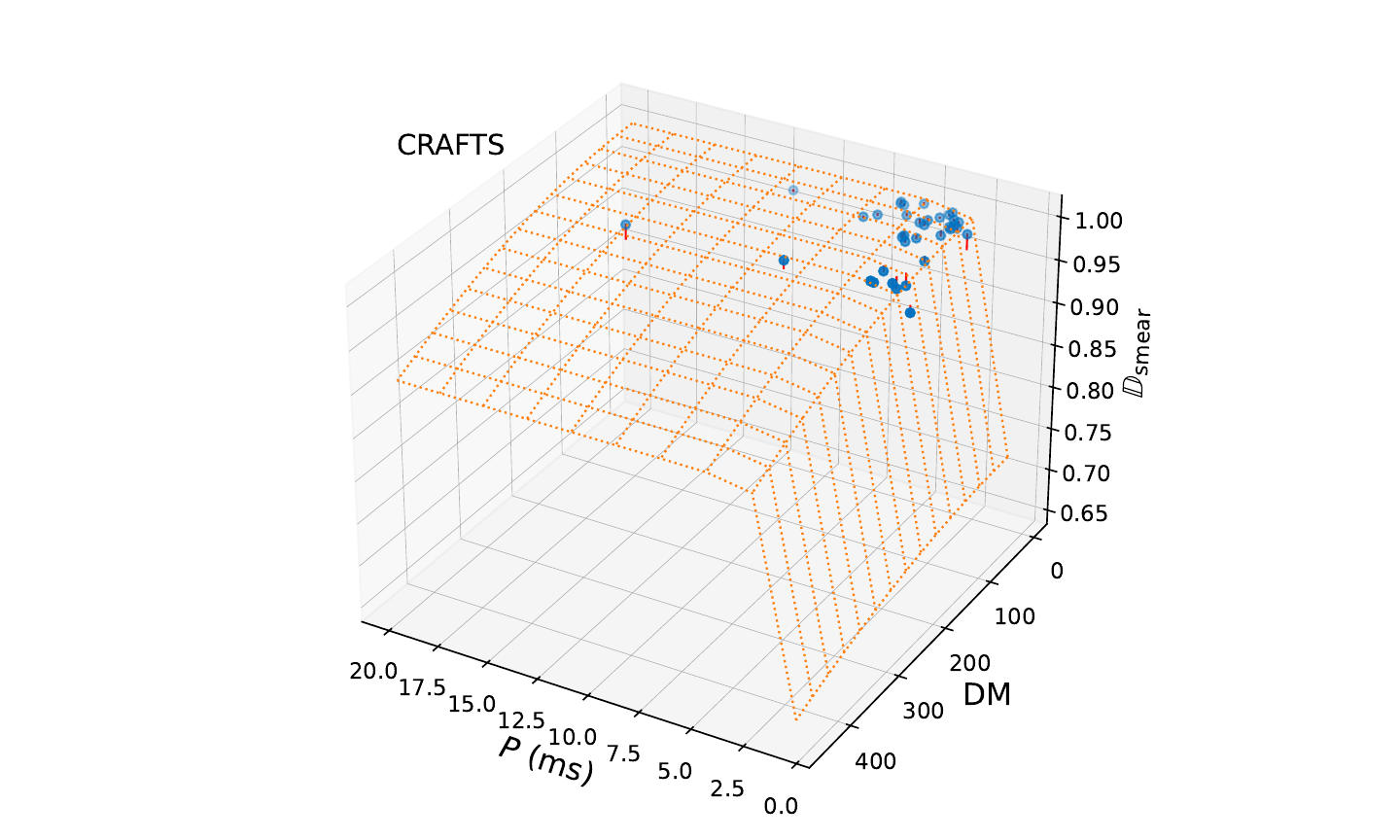}
    \includegraphics[trim={5cm 0.2cm 4.2cm 1.5cm}, clip, width=0.45\columnwidth]{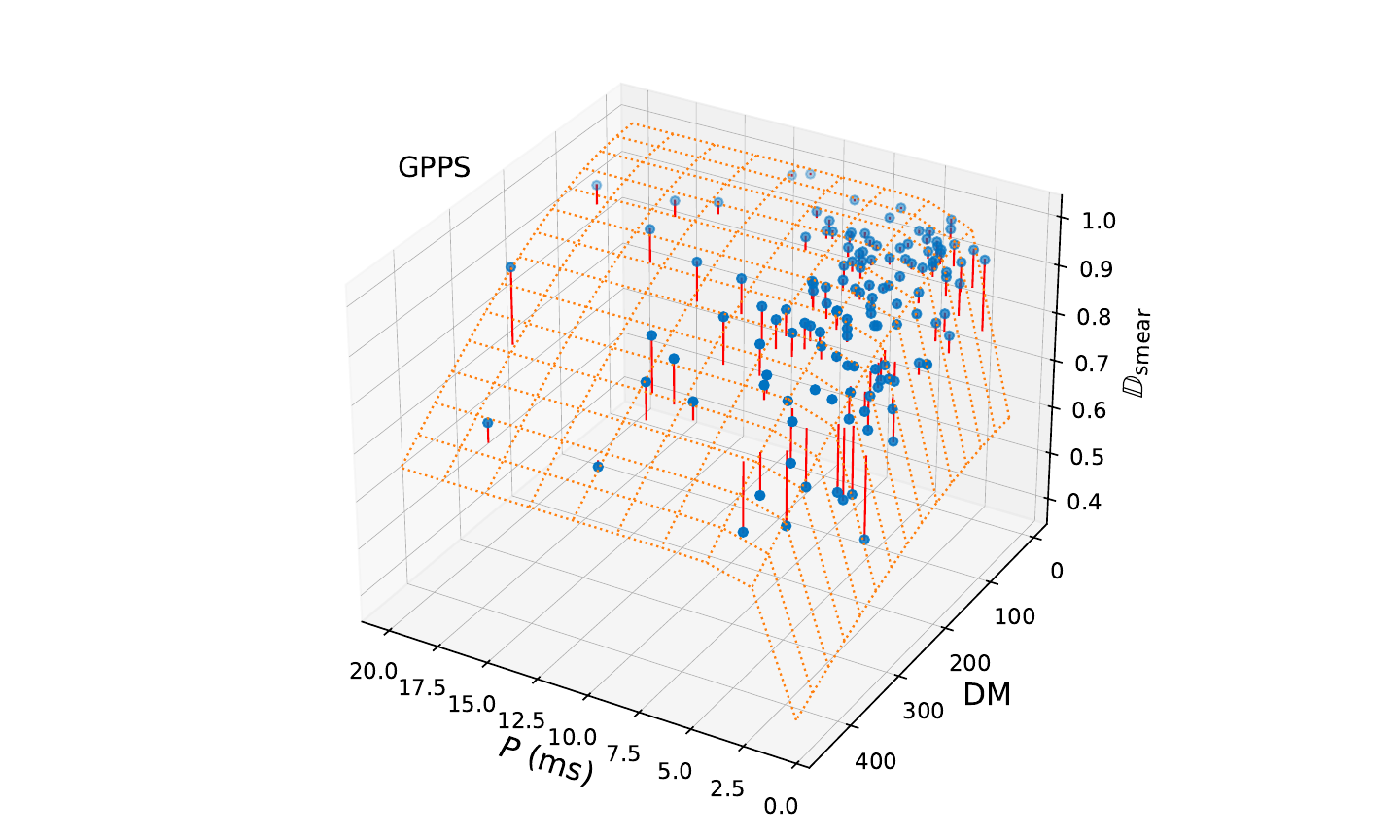}
    \includegraphics[trim={5cm 0.2cm 4.2cm 1.5cm}, clip, width=0.45\columnwidth]{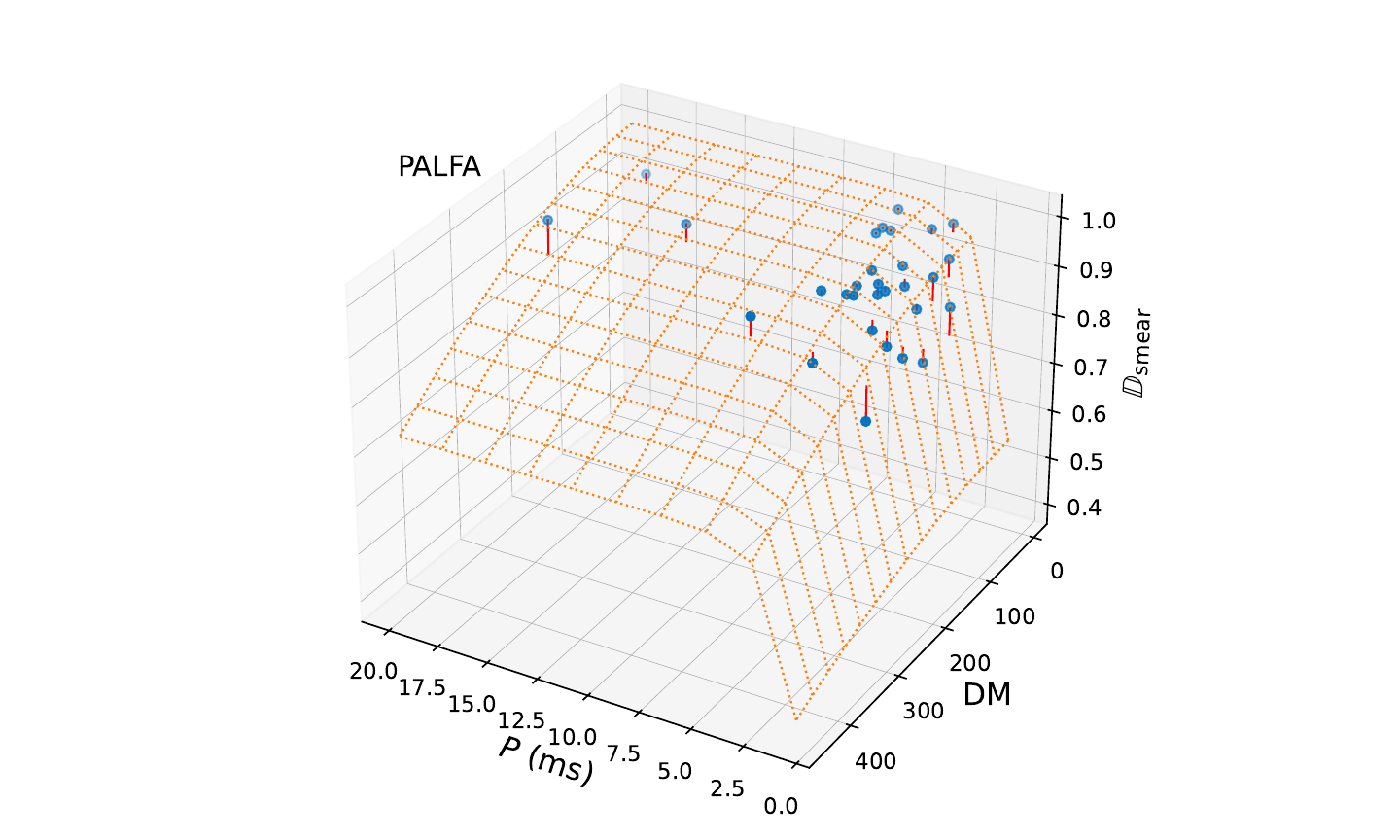}
    \caption{The detectability rate of Sample A pulsars from the five surveys and the fitting results to $p_{\rm smear}$. 
    The blue dots are the computed $\mathbb{D}_{\rm smear}$ and the yellow mesh is the fitting result with the fitting error indicated by vertical red lines.}
    \label{fig:psmear}
\end{figure}

Using the results above and Eqn.~\ref{Eqn:pSdetect}, we computed $\mathbb{D}_{\rm smear}$ for all the five surveys and show the results in Fig.~\ref{fig:psmear}. 
The selection effects are obtained by fitting $\mathbb{D}_{\rm smear}$ with $p_{\rm smear}$. 

\section{Robustness of the Bayesian method }
\label{apx:robustness}
\begin{figure}
    \centering
      \includegraphics[width=0.51\textwidth]{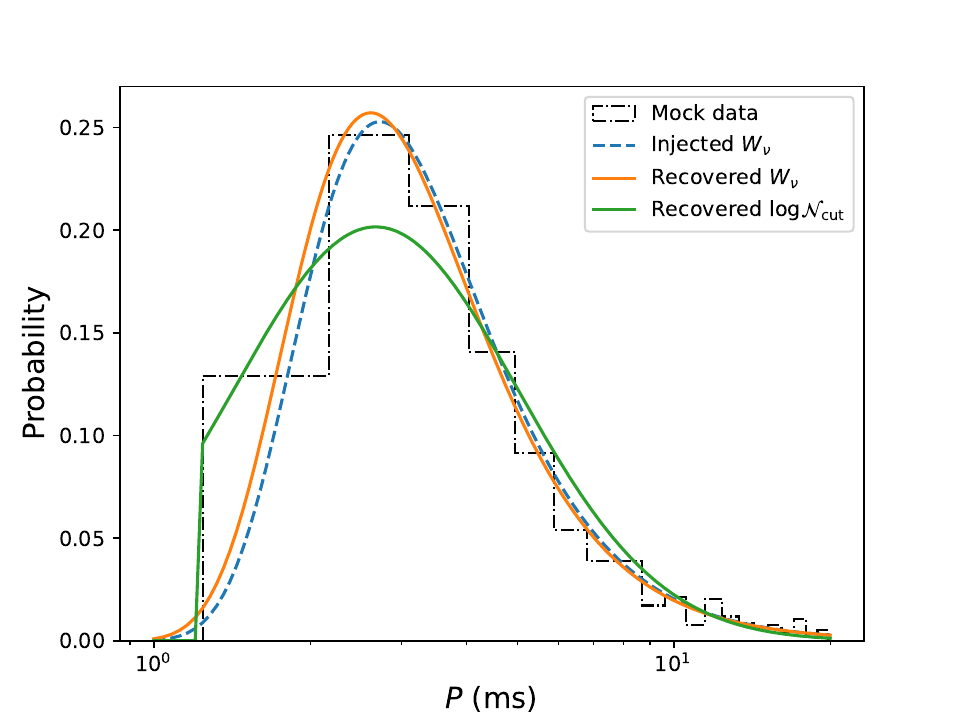}
      \includegraphics[width=0.4\textwidth]{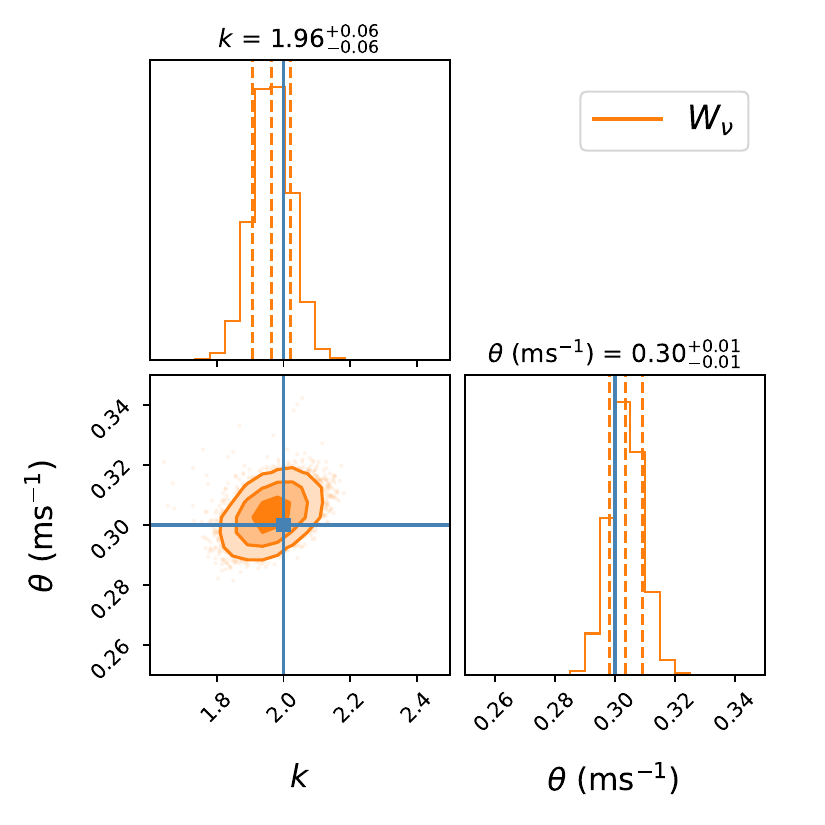}
    \caption{{\bf Left:} The mock data, the injected and reconstructed $W_\nu$ model, and the reconstructed {\tt log}$\mathcal{N}_{\rm cut}$ model. 
    {\bf Right:} Comparision between the posteriors of the $W_\nu$ model and the injected parameters used in generating the mock data.}
    \label{fig:in_re}
\end{figure}

To test the robustness of our method, we firstly generate 473 spin periods with the model $W_\nu(2, 0.3)$, simulating both the sample size and posteriors of Sample B.
For simplicity, no selection effect was considered.
The model and mock data are shown in Fig.~\ref{fig:in_re}. 
We then analyze the mock data with the $W_\nu$ and {\tt log}$\mathcal{N}_{\rm cut}$ model in the same way as the analysis of real data.
The log Bayes factor is 33.6, giving a decisive preference for $W_\nu$ over {\tt log}$\mathcal{N}_{\rm cut}$, which is clearly shown in the reconstructed results in Fig.~\ref{fig:in_re}. 
The posteriors are also consistent with the injected model parameters.
The simulations thus prove that our method is robust in both parameter estimation and model selection.

\section{Other selection effects}
\label{sec:other_sel}
Binary motion may also induce a selection bias, as the motion modulates the apparent spin frequency and reduces the chance to discover binary pulsars. 
Specifically, the frequency shift is $aT/(Pc)$, where $a$ is the acceleration, $T$ is observation length, $P$ is the spin period and $c$ is the speed of light \citep{Lor08}.
For the impact on period analysis, most recent pulsar surveys have done extensive acceleration searches and have found a lot of binary pulsars. 
The efforts thus have significantly reduced the impact, especially for binaries with wide orbits.
Actually, our analysis find no significant difference in the period distribution between isolated and binary systems (see Section~\ref{sec:sub_groups}).
While for compact binaries and for short spin periods, the effect may still be severe. 
An adequate treatment of this effect requires further investigation but is beyond the scope of this paper.

\bibliography{MSP_period}{}
\bibliographystyle{aasjournal}

\end{document}